\documentclass[useAMS,usenatbib]{mn2e}
\pdfoutput=1 
\usepackage{graphics}
\usepackage{textcomp}
\usepackage{url}

\newcommand{\note}[1]{#1} 

\title[Hydrogen volume densities in nearby galaxies I]{Hydrogen volume densities in nearby galaxies I -- \\an automated approach}
\author[J. S. Heiner, J. R. S\'{a}nchez-Gallego, L. Rousseau-Nepton and J. H. Knapen]{J. S. Heiner$^{1,4}$\thanks{E-mail: j.heiner@crya.unam.mx}, J. R. S\'{a}nchez-Gallego$^{2,3}$, L. Rousseau-Nepton$^{1}$ and J. H. Knapen$^{2,3}$\\
  $^1$D\'{e}partement de Physique, Universit\'{e} Laval, Qu\'{e}bec, QC G1V 0A6, Canada\\
  $^2$Instituto de Astrof\'{i}sica de Canarias, E-38205 La Laguna, Tenerife, Spain\\
  $^3$Departamento de Astrof\'{i}sica, Universidad de La Laguna, E-38200, La Laguna, Tenerife, Spain\\
  $^4$Centro de Radioastronom\'{i}a y Astrof\'{i}sica (CRyA), Universidad Nacional Aut\'{o}noma de M\'{e}xico, C.P. 58190 Morelia, Michoac\'{a}n, Mexico}

\begin{document}
\pagerange{\pageref{firstpage}--\pageref{lastpage}} \pubyear{2012}
\maketitle
\label{firstpage}

\begin{abstract}
Using a simple model of photodissociated atomic hydrogen on a galactic scale, it is possible to derive total hydrogen volume densities. These densities, obtained through a combination of atomic hydrogen, far-ultraviolet and metallicity data, provide an independent probe of the combined atomic and molecular hydrogen gas in galactic disks.

We present a new, flexible and fully automated procedure using this simple model. This automated method will allow us to take full advantage of a host of available data on galaxies in order to calculate total hydrogen volume densities of giant molecular clouds surrounding sites of recent star formation. So far this was only possible on a galaxy-by-galaxy basis using by-eye analysis of candidate photodissociation regions.

We test the automated method by adopting different models for the dust-to-gas ratio and comparing the resulting densities for M74, including a new metallicity map of M74 produced by integral field spectroscopy. We test the procedure against previously published M83 volume densities based on the same method and find no significant differences. The range of total hydrogen volume densities obtained for M74 is approximately 5-700 cm$^{-3}$. Different dust-to-gas ratio models do not result in measurably different densities.

The cloud densities presented here add M74 to the list of galaxies analyzed using the assumption of photodissociated atomic hydrogen occurring near sites of recent star formation and further solidify the method. For the first time, full metallicity maps were included in the analysis as opposed to metallicity gradients. The results will need to be compared to other tracers of the interstellar medium and photodissociation regions, such as CO and CII, in order to test our basic assumptions, specifically, our assumption that the HI we detect originates in photodissociation regions.
\end{abstract}


\section{Introduction}

Molecular hydrogen is a much-investigated substance, as it is generally believed to be the key ingredient to star formation.
Recent publications postulate a near-linear relation between (dense) molecular hydrogen gas densities and star formation rates \citep[e.g.][]{2011ApJ...731...25K,2011AJ....142...37S}, at least on galactic scales, fueling this view.

Yet molecular hydrogen itself is not easy to observe directly, especially outside of our own Galaxy. The closest alternative, carbonmonoxide (CO), is a tracer of dense molecular gas that is commonly used to infer molecular hydrogen column densities directly. Using $^{13}$CO is preferred \citep[][]{1978ApJS...37..407D}, but $^{12}$CO emission is stronger and therefore more readily detected. It is generally assumed that these CO intensities can be converted to molecular hydrogen column densities directly through a constant multiplication factor \citep[][]{1986A&A...154...25B}, commonly known as the X-factor. However, doubts about this conversion factor being truly constant across galaxies have been voiced since about the same time \citep[][]{1988ApJ...325..389M}. For example, \citet{1997A&A...328..471I} shows a clear metallicity dependence of the X-factor.

The 'dark molecular gas' \citep[][]{2010ApJ...716.1191W}, envelopes of molecular hydrogen without detectable CO emission, may be missed and unaccounted for in the X-factor. A recent estimate from observations of the Perseus molecular cloud in our own Galaxy by \citet{2012ApJ...748...75L} is that approximately 30 percent of the molecular gas is not accompanied by (visible) CO.
Ionized carbon [C\textsc{ii}] might be expected to trace this more diffuse molecular gas. \citet{1991ApJ...373..423S} concluded that this carbon is formed in photodissociated gas between the giant molecular cloud and the ionized gas. They claimed that CO emission originates mostly from molecular gas exposed to elevated UV fields, not from cold disk clouds. Therefore CO may trace a combination of mass and excitation.
For example, new results from the Herschel observatory in \citet{2011A&A...532A.152M} show how the CO likely gets photodissociated in the low-metallicity environment of M33. These observations of, among others, [C\textsc{ii}] are part of an ongoing effort to explain its occurrence relative to the molecular (CO) and atomic (HI) gas. Against this background, it remains important and relevant to look into and develop different ways of estimating hydrogen abundances.

An alternative approach to locate molecular clouds is to consider clusters of young, hot OB stars and the giant molecular clouds that surround them. These clouds are either the remnants of the clouds out of which these OB stars formed, or were newly formed when the gas was swept up by expanding giant HII regions. \citet{1974A&A....37...33B} describe the (ring-like) HII regions of M33, which tend to be located on the edge of HI clouds with typical shell densities of the order of 1-10 cm$^{-3}$. Atomic hydrogen disks tend to show large HI holes with diameters ranging from 40 pc to 1 kpc, where the holes tend to correlate well with OB associations \citep[M33:][]{1990A&A...229..362D}. Supershells remain actively investigated to date \citep[see e.g.][investigating supershells in our own Galaxy]{2011ApJ...741...85D}. 

In this context, the atomic hydrogen associated to the OB star clusters should be a direct product of photodissociation. Certainly the OB stars surrounded by HI supershells conjure up an image of gas being blown into bubbles with molecular clouds forming in the shells, to be partially photodissociated by the UV radiation of the young stars. For a general review of photodissociation regions (PDRs), see \citet{1999RvMP...71..173H}. In M83, atomic hydrogen appears to be photodissociated on a large scale \citep[][]{1986Natur.319..296A}. This property can be used to infer the total hydrogen gas density including the molecular hydrogen gas. For a recent, galaxy-wide application of this view of PDR-produced HI, see \citet{2011ApJ...739...97H}.

This procedure of locating potential PDR sites around OB star clusters, that we refer to as the 'PDR method', was presented previously for M33, M81 and M83 \citep[][]{1997ApJ...487..171A,2008ApJ...673..798H,2008A&A...489..533H,2011MNRAS.416....2H}; see also \citet{2000ApJ...538..608S} for M101. The method was investigated for selection effects, resolution and distance effects. Until now, the analysis was carried out mostly by hand. This makes the procedure slow to apply, and also potentially introduces biases. 
Therefore, we want to take the first step towards applying this method more routinely to a large number of galaxies and make the method widely available. To this end we present an automated 'pipeline' to extract total hydrogen volume density measurements from a far-ultraviolet image, an atomic hydrogen image, and a dust model or dust map of a galaxy. These spot measurements present total hydrogen gas densities at the location of potential PDRs across galactic disks. The densities are independent of CO measurements and independent of any assumptions about the presence of molecular gas. They also do not depend on measurements (or the presence) of warm dust. We prefer this method because of the straightforward physics behind it, namely the process of photodissociation by ultraviolet radiation regulated by the local metallicity. 

For this first paper presenting the automated 'pipeline', we selected M74 as our first target, because it is practically face-on, reducing the influence of projection effects. Another reason to select M74 is the availability of detailed metallicity data, useful to increase the accuracy of the PDR method. This is the first time a metallicity map has been used in the method. M74 is more distant than M33, M81 or M83, but the tell-tale combination of far-UV sources surrounded by patches of HI emission is present. Paper II will feature a larger sample of nearby galaxies analyzed with our PDR method, enabling a more global evaluation of the method and its results.

This paper is arranged as follows. First we briefly review the basic PDR theory, followed by the presentation of our automated measurement method. Then we present a newly obtained metallicity map of M74 that we use to derive the dust-to-gas ratio. We give the results of the PDR method for M74 using various metallicity measurements and compare the total hydrogen volume densities to those previously obtained in M83 \citep[][hereafter H08b]{2008A&A...489..533H}. 

\section{Deriving total hydrogen densities from atomic hydrogen}
\label{sec:pdrmethodtheory}

This work is based on the assumption that a large part of atomic hydrogen in the disks of spiral galaxies is produced in photodissociation regions, as opposed to being 'primordial' and predating recent episodes of star formation. 
We briefly summarize this procedure, that comprises the PDR method.

First, we need to locate so-called 'candidate PDRs', regions in galaxies dominated by clusters of young, hot OB stars. We expect these clusters to be surrounded by remnants of their parent giant molecular clouds, or alternatively, by recently formed GMCs in the expanding boundary of giant extragalactic HII regions (GEHRs) with OB stars at the center. These regions, with a size of up to several hundred parsec, we collectively call candidate PDRs. Under the influence of the far-ultraviolet radiation from the central cluster of stars, photodissociated atomic hydrogen appears on the surface of nearby GMCs.

This PDR-produced HI is identified as locally enhanced HI column densities with a general HI background component. In general, several of these HI patches can be found per candidate PDR. Assuming a spherical geometry, we can then calculate the incident UV flux $G_0$ on these HI patches. Together with the local dust-to-gas ratio (DGR) we can then calculate the balance of photodissociation and therefore the total hydrogen volume density at the position of the HI column.

The final equation describing the balance of photodissociation is derived from \citet{1988ApJ...332..400S} and \citet{2004ASSL..319..731A}, using a simple, two-level molecular hydrogen, one-dimensional approach. This allows us to calculate the total hydrogen volume density from the local atomic hydrogen column density, the incident UV flux and the local dust-to-gas ratio.  It includes recent improvements suggested by \citet{2009ApJ...694..978H}. The result is
\begin{equation}
  n = 106 \frac{G_0}{\left(\delta/\delta_0\right)^{0.3}}\left[\exp{\left(\frac{N_{HI}(\delta/\delta_0)}{7.8 \times 10^{20}}\right)}-1\right]^{-1} \left[\rm{cm^{-3}}\right],
  \label{eqn:ntot}
\end{equation}
where the factor $\left(\delta/\delta_0\right)^{-0.3}$ is just slightly different from the factor $\left(\delta/\delta_0\right)^{-0.5}$ used previously \citep[e.g.][]{2011MNRAS.416....2H}. $\delta/\delta_0$ is the dust-to-gas ratio (scaled to the solar neighborhood value). Here, $n$ is the total hydrogen volume density. $G_0$ is the incident far-UV flux scaled to the solar neighborhood value, and $N_{HI}$ is the neutral hydrogen column density in $\rm{cm}^{-2}$. In the context of a simple spherical geometry, the measured HI column density can be assumed to be equivalent to the HI column from central OB source in the direction of the GMCs surrounding it.

The difference between the equation used in earlier applications of this method and the current one is below the estimated uncertainty for normal values of $\delta/\delta_0$. 

\section{Automated implementation of the PDR method}
\label{sec:pipeline}

The purpose behind developing the pipeline we present here is to provide the astrophysical community with a transparent, yet configurable tool to apply the PDR method.

The way the pipeline works can be roughly divided into three main steps. Firstly (Section \ref{sec:SourceSelection}), candidate PDRs are identified. Secondly, (Section \ref{sec:Measurements}), measurements are taken from each candidate PDR (fluxes, geometry, column densities and dust-to-gas ratios). Finally, total hydrogen volume densities are computed using the simple photodissociation model described in Section \ref{sec:pdrmethodtheory}.

\subsection{Source selection}
\label{sec:SourceSelection}

Our method uses regions of strong FUV emission to select candidate PDRs. We use the clumpfind algorithm of \citet{1994ApJ...428..693W} on FUV imagery to find and mask regions of enhanced emission. Clumpfind works by calculating contour levels on the selected image. The pixels contained within each contour level are labeled as belonging to the same region. When a new contour level is calculated, the already defined regions lying within the new region are joined to the latter. Similarly, when two or more regions become adjacent, all of them are incorporated into the largest one. This process continues until each pixel belongs to one region. The final product of clumpfind is a new file in which each pixel of the original image is labeled with the number of the region it belongs to. Hereafter, we will call this file a mask.

Unless told otherwise, clumpfind will use all the pixels in the image. In practice, a certain range of values must be excluded to avoid including the diffuse FUV background emission. We define this threshold as a function of the average noise background level, $\sigma_{\rm N}$, which can be defined manually as an input parameter or interactively calculated. In the latter case, the user is asked to select a region of uniform FUV emission (i.e., neither a strong peak nor a region completely free of FUV emission). $\sigma_{\rm N}$ is then estimated through and iterative process: Firstly, the median and standard deviation, $\sigma$, of all the pixels in the region is calculated; secondly, those pixels with values above or below 3$\sigma$ are rejected (thus removing potential cosmic rays or ``hot'' pixels); this procedure is repeated until the median value changes by less than 1\%. Once $\sigma_{\rm N}$ is known, the lowest contour level is defined as a number of times the background noise ($n$). While choosing an $n$ too low (for example, $n=1$) can ensure that no sources of emission are overlooked, spurious regions of background emission will likely be included. On the other hand, too strict a threshold value ($n>5$) may miss some of the weakest regions, thus biasing the sample of HII regions. Another important parameter is the number of contours to fit. We have found that a number too large (for example 50 contour levels) increases the computation time while not yielding any improvement (in some cases even producing worse results). The optimal value will depend on the galaxy but our experience indicates that 20 contour levels (the default value for the pipeline) produces good results without dramatically increase computation time.

{Our tests using different galaxies indicate that the results of clumpfind are fairly robust against different sets of parameters, with the number of contours being the most critical value to obtain good results over a whole galaxy. Depending on the chosen background level, far-UV clumps are identified at a larger galactic radius at a cost of potentially spurious detections. A balance must be struck by the user, who will need to assess whether any obvious sources of UV emission were missed and whether this can be remedied by changing the number of contours and the background level. We do not aim to extract a complete sample of candidate PDRs, but rather a representative sample in terms of far-UV flux and spread in galactocentric radius. The purpose of this method is to produce a statistically significant catalog of candidate PDRs in a way that minimizes the involvement of the user in the process.}

{The initial mask produced by clumpfind has two issues that need to be dealt with. Firstly, some regions are broken into several parts that are still part of the same OB cluster. Secondly, even if regions are separate clusters they can be close enough to cause confusion. These issues} can be solved by applying the following rejection criteria. First, we consider the typical size of a GMC and, if two regions are closer than a certain value (typically $\approx500\,\rm pc$), the weakest one is rejected or joined to the strongest (the behavior can be defined as an input parameter). The ``strength'' of a region can be determined from its estimated total flux or the value of the peak pixel (again, configurable via an input parameter). Similarly, when two regions are adjacent, a contrast test is carried out with the weakest region being rejected or joined to the strongest one. We find that this method allows us to resolve the great majority of the regions identified by clumpfind. This provides a selection of candidate PDRs similar to the more or less 'by-eye' method of selecting them used in our previous publications (like H08b).

{Finally, the user may opt to bypass the source detection with clumpfind altogether and directly supply a list of coordinates. These coordinates could be, for example, a list of HII regions or a hand-picked list. This also means that the list of regions found by the clumpfind algorithm can be edited by hand if so desired. In this manner two use cases are distinguished that can be mixed, namely a 'blind' selection of candidate PDRs and a pre-defined sample set of objects.}

Once the final FUV and HI regions have been selected, the pipeline produces catalogs in both VOTable\footnote{\url{http://www.ivoa.net/cgi-bin/twiki/bin/view/IVOA/IvoaVOTable}} and DS9\footnote{\url{http://hea-www.harvard.edu/RD/ds9/}} formats. FITS files containing each individual region in the far-UV and 21-cm are also created. {The DS9 region file in particular facilitates easy inspection of the coordinates of the candidate PDRs, which is highly recommended considering the nature of the previously outlined selection process.}

\subsection{Measurements and calculations}
\label{sec:Measurements}

{Far-ultraviolet fluxes are measured using the task \verb+phot+  from  the package \verb+digiphot+ in IRAF\footnote{IRAF is distributed by the National Optical Astronomy Observatory, which is operated by the Association of Universities for Research in Astronomy (AURA) under cooperative agreement with the National Science Foundation.}. For each region, \verb+phot+ determines the peak position of the candidate PDR and performs a series of concentric aperture photometries. Average fluxes per ring are calculated for each radius and the optimal aperture is determined where the average flux reaches its first local minimum after it has dropped to (by default, but configurable) below 50\% the peak flux. Regions for which the optimal aperture cannot be determined using this method are rejected, which means that sources confused over larger areas and sources with a poor source contrast are excluded. The local background level is calculated using the average flux of the concentric aperture at the first local minimum.}

{Based on the optimal aperture and local background emission, the pipeline calculates the net flux in the region as the cumulative flux at the given optimal aperture minus the estimated background emission for that area. The relative error $\sigma_{\rm FUV}$ is defined as a fraction of the net flux, set as a parameter in the configuration file.}

Formally, we consider local peaks in the HI column density close to clusters of OB stars as PDR-produced HI patches. The column densities of these patches can be used directly in our calculations. We use SExtractor \citep[][]{1996A&AS..117..393B} to detect these patches automatically. While SExtractor was developed for star/galaxy recognition, we find it quite suitable for locating HI patches. {In order to avoid confusion with larger scale structures, SExtractior is applied for each region on a subset of the HI map defined around the FUV peak.} The SExtractor results are easily read and incorporated using Harry Ferguson's sextutils \footnote{\url{http://www.stsci.edu/~ferguson/software/pygoodsdist/pygoods/sextutils.py}}.

{The HI patches are corrected for diffuse emission using an overall background level which needs to be specified as an input parameter. The uncertainty $\sigma_{\rm HI}$ is set to half the HI background level.}

Using the recorded locations of each HI patch combined with the locations of the central UV sources, the source-patch separations $\rho_{HI}$ are calculated on the plane of the sky. With $\rho_{HI}$ we can calculate the incident flux $G_0$ and the source contrast at each HI patch, correcting for a (fixed) foreground extinction. The dust-to-gas ratio is determined at the location of the HI patch. {The pipeline provides three options for this: Using a simple galaxy-wide constant value, or a metallicity slope and a corresponding galactocentric radius for each patch, or an oxygen abundance map where a value can be read directly from the map. This last option in particular is expected to become more useful with the increasing availability of maps obtained from integral field spectroscopy.} Finally, the total hydrogen volume density at each HI patch is calculated using Equation \ref{eqn:ntot}. We select the results to have a source contrast of at least unity.

\subsection{Configuration file}
\label{sec:ConfigurationFile}

One of the main goals we pursued while developing this automated method was to make sure that the tool was highly customizable. Although we strived to provide reasonable default values, most of the pipeline control parameters can be adjusted via a configuration file. {The most important of} these parameters are the number of levels used by clumpfind, the background level of the atomic hydrogen map and the minimum acceptable signal-to-noise {and} several physical parameters which control the FUV rejection mechanism. {By providing sensible default parameters we aim to ensure that acceptable results are obtained with the user mainly providing the input maps. The parameters mainly influence the localization of candidate PDRs and HI patches (through SExtractor) but not the actual physics that happens in the pipeline.}
 
\subsection{Obtaining the software}

The pipeline is implemented using the Python programming language\footnote{\url{http://www.python.org}}. The sofware will be released officially with Paper II (S\'{a}nchez-Gallego et al., in prep.), but in the meantime it can be obtained from the authors directly. 
The authors would be grateful to receive any feedback or suggestions that help to improve future versions of the tool.

\section{Data}

\subsection{Basic data and archival images of M74}

The local radiation field impinging on our candidate PDRs is estimated from far-UV photometry of the publicly available GALEX image of M74 \citep[][]{2007ApJS..173..185G}. We identify atomic hydrogen features (that we expect to be PDR-produced) from the THINGS HI image \citep[][]{2008AJ....136.2563W}. The GALEX image has an angular resolution of approximately 4 arcsec, whereas the THINGS image has a resolution of 6 arcsec (using the image produced with robust weighting). We list basic properties of M74 in Table \ref{tab:properties}. The adopted distance to M74 influences the deprojected sizes of our candidate PDRs. It also affects the value of $R_{25}$ in units of kpc, but this has no influence on our results since it is just a scale of galactocentric radius.
We chose this distance for consistency with \citet{2010ApJ...714..571W}, who in turn adopted the distance measurement from the catalog of neighboring galaxies by \citet{2004AJ....127.2031K}. The foreground FUV extinction is derived from the visual extinction E(B-V) using the expression from \citet{2007ApJS..173..185G} and used to correct the measured GALEX fluxes. It is assumed that the candidate PDRs have an approximately spherical geometry, in which internal extinction between the central UV source and the HI patches is equal to the internal extinction towards the observer. In this scenario the internal extinction corrections cancel out and will be 0. If, due to projection effects or because the spherical assumption is inaccurate, the extinction is non-zero, we will underestimate the incident UV flux and as a consequence underestimate the hydrogen density as well. We adopt a 'plausibility limit' of 500 pc, beyond which HI patches are assumed not to be associated with the central UV source. This limit compares favorably to a neutral gas disk scale height of 700 pc as adopted by \citet{2010ApJ...715..902H} in their study of the outer disk of M74.

\subsection{Metallicity}

The balance of photodissociation is very sensitive to the local dust-to-gas ratio (sometimes abbreviated as DGR). We use the metallicity 12 + log(O/H) to derive the dust-to-gas ratio scaled to the solar neighborhood value $\delta/\delta_0$. After \citet{1990A&A...236..237I}, we assume a linear relation between the two \citep[more recently, see][]{2011A&A...532A..56G}. After adopting a solar oxygen abundance of 8.69 \citep{2001ApJ...556L..63A} a dust-to-gas ratio can be derived.
Typically, metallicity gradients are available in the literature for spiral galaxies. Since the scatter in such measurements is considerable, and likely intrinsic \citep{2008ApJ...675.1213R}, local metallicity values are preferred. In the case of M74, we present a new metallicity map that we used to improve the total hydrogen density measurements. We also use the metallicity map of M74 published by \citet{2011MNRAS.410..313S} and their derived metallicity gradient to compare the results.


\subsubsection{Metallicity slope}

Assuming a constant slope in the metallicity of M74 as a function of galactocentric radius, we can obtain a value of the oxygen abundance (and by extension the dust-to-gas ratio) for every candidate PDR. The most recent result for M74 comes from \citet{2011MNRAS.415.2439R}, based on their M74 metallicity map. Of the various methods they describe, we adopt the one using the O3N2 index from \citet{2004MNRAS.348L..59P}:
\begin{equation}
  \log(O/H) / dex~\rho_{25} = -0.66,
  \label{eqn:OHarcmin}
\end{equation}
where $\rho_{25}$ is the optical radius of M74 (in arcmin).

After correcting for the different distance to M74 and using a solar oxygen abundance of 8.69 we arrive at 
\begin{equation}
  \mathbf{\log}(\delta/\delta_0) = -0.060 \mathbf{\times} R + 0.24,
  \label{eqn:dd0slope}
\end{equation}
with R in kpc. By adopting the O3N2 result we maintain consistency with the metallicity map we present here that is derived using the strong line relations presented in the same paper.

\subsubsection{A new metallicity map of M74}

The M74 metallicity map presented in this paper was produced using a data cube obtained in September 2008 with the imaging Fourier transform spectrometer SpIOMM \citep[][]{2003SPIE.4842..392G,2008SPIE.7014E.246D,2008SPIE.7014E.245B,2010AJ....139.2083C} attached to the 1.6 m telescope of the Observatoire du Mont-M\'{e}gantic. A red filter, covering the wavelength range from 6480 to 6820 \AA\ and a folding order of 16 with 206 steps was used for a spectral resolution of \note{around 5} \AA\ and a total exposure time of 3 hours 15 minutes.

The SpIOMM data cube is composed of interferograms. An image of the galaxy is seen on each of them, and as a whole, they allow us to get a spectrum for each pixel of the image. The detector used was a Princeton Instrument EEV CCD camera with 1340 $\times$ 1300 pixels with a 12\note{$^\prime$} circular field of view. The data were binned 2 $\times$ 2 during the readout in order to match the size of a pixel (1.1$^{\prime\prime}$ when binned) with the seeing (of the order of 1.5$^{\prime\prime}$). This gave us a total of 435500 spectra.

The first step of the data reduction was to align the interferograms using the centroids of field stars. Then, for each interferogram, we subtracted a bias and a dark image (collected with the same instrument).  We used a dome flat image (with the instrument mobile mirror fixed away from the zero path difference position) to calibrate the pixel-by-pixel response of the detector. Cosmic rays were subtracted in each interferogram. The sky background variation was corrected by considering an image region away from the galaxy and devoid of other objects. In this region, we measured the average flux value between two consecutive interferograms (one with a constructive and one with a destructive interference). All consecutive interferograms should have presented the same average sky value, but because of changing conditions during the long exposure, this was not the case. Therefore the interferograms were corrected in intensity to take into account the measured variation seen in the sky. \note{It should be noted that as of 2011, SpIOMM has been equipped with a second CCD camera, drastically improving sensitivity and permitting measuring the total flux values. This allows corrections for sky variation to be made during the observations.} Finally, the Fourier transform was applied to the data cube and the wavelength calibration was done using a high-resolution data cube of a reference HeNe laser at 6328 \AA. All these operations were performed with IDL (the Interactive Data Language).

Line measurements for the HII regions should be done on the average spectrum for each regions in order to increase the signal-to-noise ratio. Therefore, HII regions have been identified  using the {\tt HIIphot} program of \citet{2000AJ....120.3070T}, created specifically for this purpose. Using the SpIOMM data, we first created an H$\alpha$ map, a red continuum map, and a combined H$\alpha$+continuum map of the galaxy, where {\tt HIIphot} calculated HII region contours. The relevant parameters we used for {\tt HIIphot} were: {\tt NIIpercent} = 1\% (percentage of [NII] contamination in the H$\alpha$ flux) and {\tt size\_max} = 500 pc (maximum size of the HII regions). A total of 541 HII regions have been identified. They range in size from 4 to 93 pixels. An example interferogram is shown in Figure \ref{fig:interferogram}. At this point, we did not directly combine the spectra included in each region, but reapplied the Fourier transform to the data cube after combining the pixels corresponding to the HII regions in each interferogram. This step also improved the signal-to-noise ratio in the final spectrum for each HII region. The sky background level and lines are removed by building an average sky spectrum using all pixels away from the galaxy and other objects. An example HII region spectrum is shown in Figure \ref{fig:spectrum}.

Since the goal here was to obtain a  metallicity map from measurements of the two nearby lines of H$\alpha$ and [NII]$\lambda$6583, no flux calibration and extinction corrections have been done yet (a complete analysis of these data, along with a blue cube, will be part of another paper by Rousseau-Nepton). The two lines have been measured using a Gaussian fit with the IDL task {\tt gaussfit}. The uncertainties for  these measurements are dominated by the position of the continuum.

\begin{figure}
  \resizebox{\columnwidth}{!}{\includegraphics{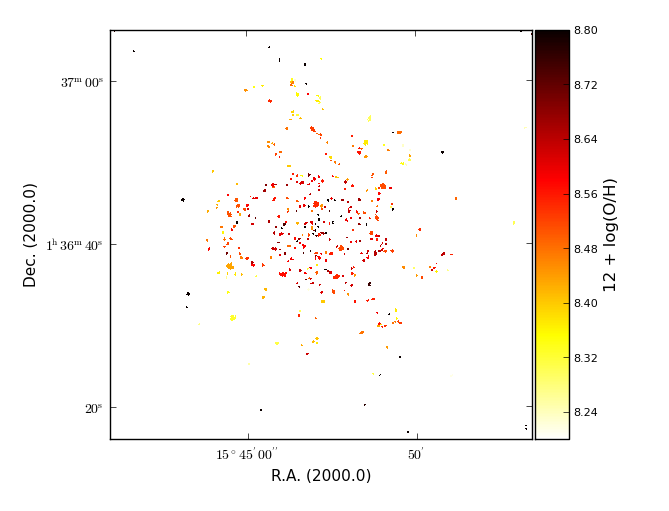}}
  \caption{\label{fig:metalmap} Map of the 12 + log(O/H) values of the HII regions obtained from the SpIOMM observations. A color version of this map is available in the electronic version of this journal.}
\end{figure}

\begin{figure}
  \resizebox{\columnwidth}{!}{\includegraphics{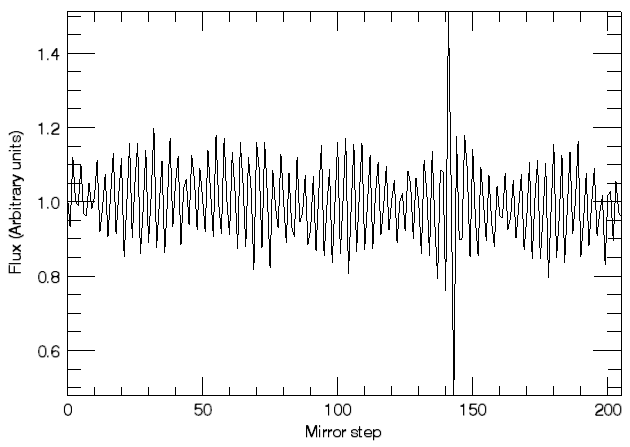}}
  \caption{\label{fig:interferogram} Sample interferogram of an HII region in M74.}
\end{figure}

\begin{figure}
  \resizebox{\columnwidth}{!}{\includegraphics{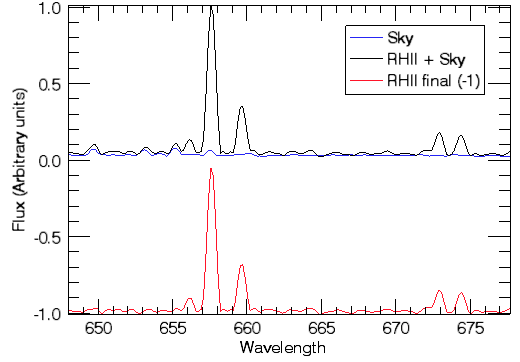}}
  \caption{\label{fig:spectrum} Sample HII region spectrum showing the H$\alpha$ and [NII] lines as well as the SII doublet (not used here). The sky-subtracted spectrum is offset for clarity.}
\end{figure}

The metallicity (Z) was finally calculated using the [NII]/H$\alpha$ ratio and the polynomial relations of \citet{2004MNRAS.348L..59P}. Altough this ratio gives a good estimate of the metallicity, it is sensitive to the ionization factor (q). The degeneracy effect is even more important at high Z.  We calculated the error due to this degeneracy effect using the relations between the [NII]/H$\alpha$ ratio and the Z, q value that can be found in \citet{2002ApJS..142...35K}. Figure \ref{fig:metalmap} shows our metallicity map. 

We compare our metallicity map to the one obtained with the PPAK IFS Nearby Galaxies Survey (PINGS), as presented in \citet{2011MNRAS.410..313S} and \citet{2011MNRAS.415.2439R}. There are a number of differences. While both the SpIOMM and PPAK instruments are integral field spectrographs, they are constructed quite differently. SpIOMM is a fourier transform imaging spectrograph with a field of view large enough to capture the entire optical extent of M74, but at a cost of wavelength coverage. PPAK is a fiber spectrograph with a comparatively limited field of view and a mosaic of M74 was constructed with it, covering a smaller area than SpIOMM was able to cover. PPAK's higher sensitivity \note{(mostly due to it being mounted on a bigger telescope)} and wavelength coverage allowed various ways of analyzing the results. In this paper, we simply use the resulting oxygen abundance 12 + log(O/H) map, with 
\begin{equation}
  \mathbf{\log(\delta/\delta_0)} = (12+\log(O/H)) - 8.69.
  \label{eqn:dd0map}
\end{equation}

A detailed comparison between the two different metallicity maps is beyond the scope of this paper. The full SpIOMM M74 metallicity map and comparisons with other published data will be presented in a different publication (Rousseau-Nepton et al., in prep.).

\section{Hydrogen densities in M74 candidate PDRs}

Here we present total hydrogen volume densities calculated as in Eq. \ref{eqn:ntot} near candidate PDRs across the disk of M74 using different sources for the local dust-to-gas ratio. 

\begin{figure*}
  \resizebox{\columnwidth}{!}{\includegraphics{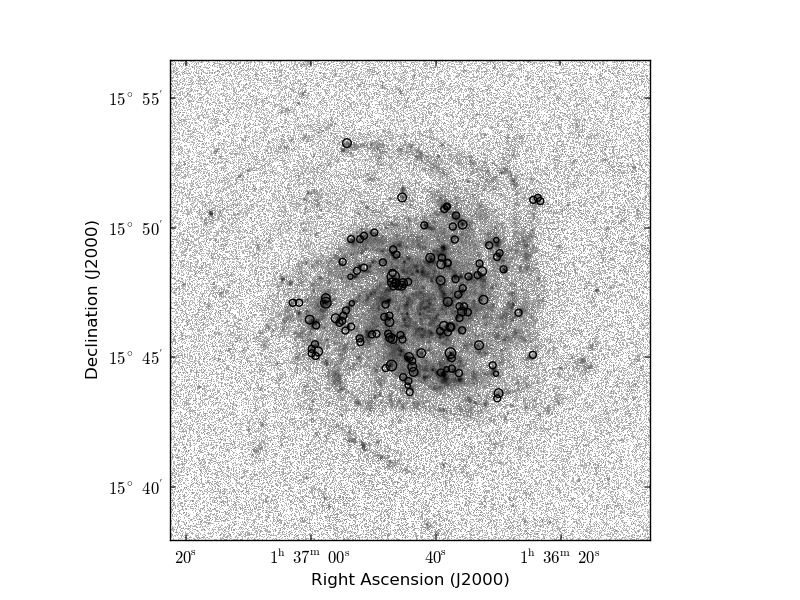}}
  \caption{\label{fig:n628_locplot} Total hydrogen volume densities across the disk of M74 near candidate PDRs as identified by clumpfind plotted on top of the GALEX far-UV image. The radius of the circles corresponds to the rounded logarithmic value of $n_{tot}$.}
\end{figure*}

\begin{figure}
  \resizebox{\columnwidth}{!}{\includegraphics{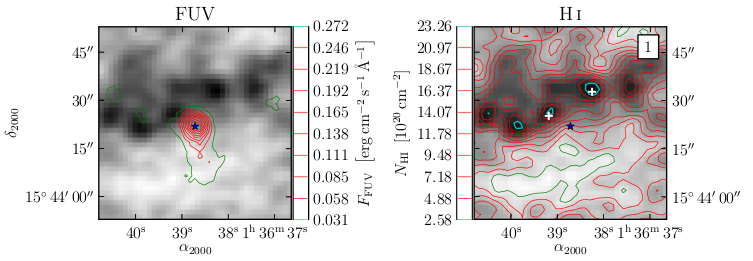}}
  \caption{\label{fig:n628_cont1} Candidate PDR in M74. On the left hand side, contours mark the far-UV flux. On the right hand side, contours mark the HI column density. Both panels have the HI column density plotted in grayscale. The center of the region is marked with a star. Selected HI patches are marked with a white plus-sign. Figures for all M74 candidate PDRs are available in the electronic version of this journal.}
\end{figure}

The candidate PDRs are located in the manner described in Section \ref{sec:pipeline}. The selected regions are shown in Figure \ref{fig:n628_locplot} (the circles indicate the HI patches surrounding the candidate PDRs). We also created far-UV/HI overlay plots. An example is shown in Figure \ref{fig:n628_cont1}. We note that the regions found by clumpfind are all inside the optical radius $R_{25}$ due to our choice of a detection threshold, although M74 is known to have a young stellar population out to two times this radius \citep[][]{2007AJ....134..135Z}. Additionally, the metallicity maps are limited to the optical radius as well, which is the focus of this work.

\begin{figure}
  \resizebox{\columnwidth}{!}{\includegraphics{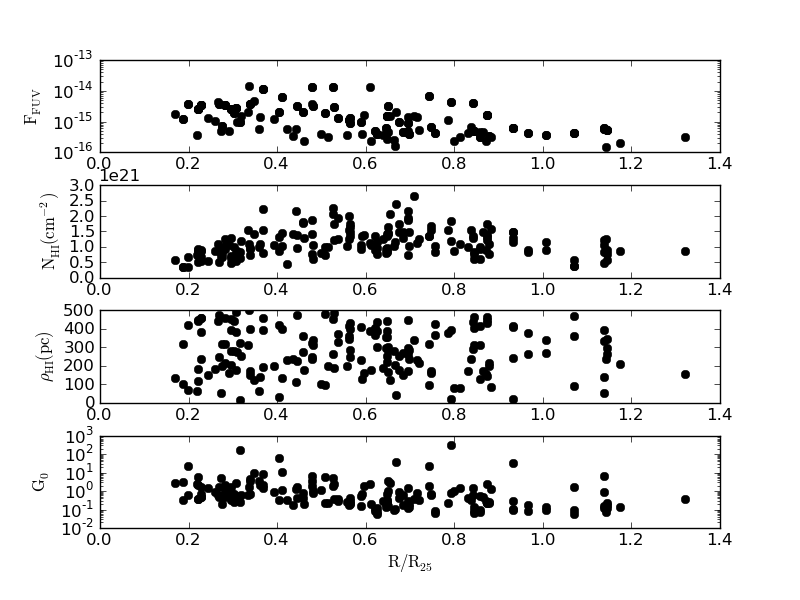}}
  \caption{\label{fig:panel_FNrG} From top to bottom: M74 measured far-UV fluxes, HI column densities, separation between the central UV source and the candidate PDRs and incident fluxes $G_0$. Fluxes are in units of ergs cm$^{-2}$ s$^{-1}$ \AA$^{-1}$.}
\end{figure}

Fig. \ref{fig:panel_FNrG} shows from top to bottom the measurements obtained from our pipeline: the far-UV fluxes of candidate PDRs across M74, the associated HI column densities found near the UV sources, the separation between the central UV sources and the associated HI patches and the resulting incident UV fluxes $G_0$ at each HI patch. The results were selected to have a source contrast $G_0/G_{bg}$ of at least 1. 

The full dataset is available electronically, but an example is shown in Table \ref{tab:results}. The datasets are in VOtable format, and the candidate PDRs are numbered consistently in the different M74 files of the densities derived from the metallicity slope, PINGS metallicities and SpIOMM metallicities. Not all candidate PDRs located through clumpfind result in reported total hydrogen volume densities, since they are dropped if the source contrast is too low, no HI patches are found nearby or if no metallicity data is available at the source location. For the PINGS map we did not have individual uncertainties so they are not included. The uncertainties in the metallicity slope values are estimated conservatively at 10\%.

The measured HI column densities increase steadily with galactocentric radius before flattening out, although the exact trend is unclear. The HI column densities peak near $3 \times 10^{21}~\rm{cm}^{-2}$, reflecting perhaps a beam smoothing issue similar to the one found in HI column densities obtained from M83 in H08b. A constant HI background level of $2.9 \times 10^{20}~\rm{cm}^{-2}$ was subtracted, which has very little influence on the final total hydrogen volume density.

The distribution of the values of $\rho_{HI}$ is not as discrete as in our previous work \citep[e.g. Fig. 4 in][]{2011MNRAS.416....2H} due to a different measurement technique. Instead of manually locating HI patches within increasing annuli at discrete separations, we have SExtractor locate the HI patches and use their exact location to calculate the separation, deprojected using the galaxy's position angle and declination as before. The accuracy of the measurement of this separation is still limited by the angular resolution of the HI map and, more importantly, by the unknown 3-dimensional position of the HI patches with respect to the central UV source. 

The values of $G_0$ do not show a strong radial trend, which is consistent with previous observations of other galaxies.

\begin{figure}
  \resizebox{\columnwidth}{!}{\includegraphics{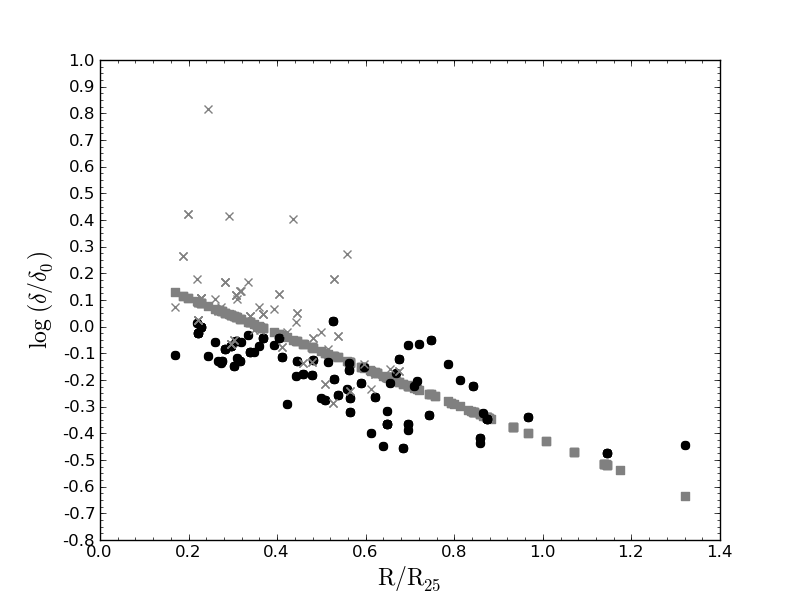}}
  \caption{\label{fig:metallicities} Dust-to-gas ratios at each candidate PDR. The black circles are the values taken from the SpIOMM metallicity map, the values computed from a fixed metallicity slope are gray squares and the values taken from the \citet{2011MNRAS.415.2439R} metallicity map are represented by gray crosses.}
\end{figure}

We compared three different sources of the dust-to-gas ratio (see Fig. \ref{fig:metallicities}): the SpIOMM metallicity map (black dots), and the metallicity map and gradient from \citet{2011MNRAS.415.2439R}. The gradient is based on their metallicity map, so it follows the actual range of metallicities. It can be seen that the metallicities from the SpIOMM metallicity map are systematically lower across most of the range of galactocentric radii.

\begin{figure}
  \resizebox{\columnwidth}{!}{\includegraphics{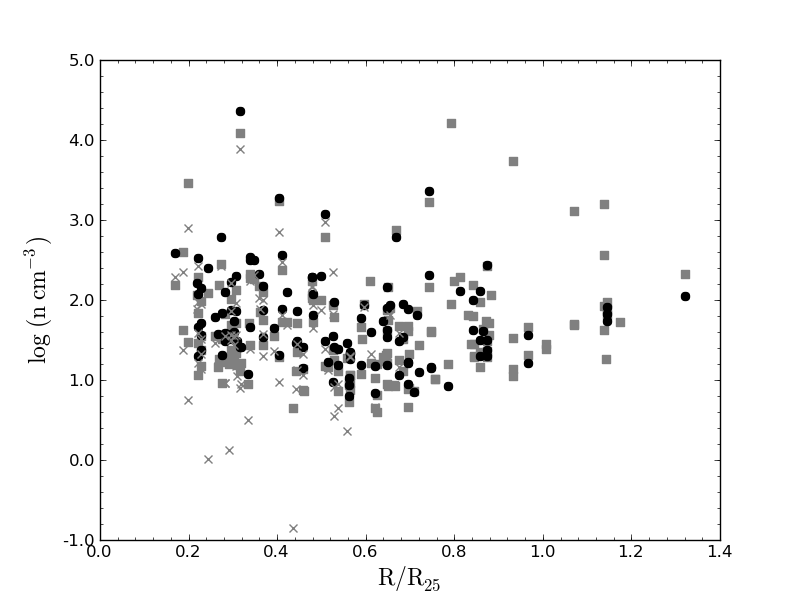}}
  \caption{\label{fig:m74_ntot} Total hydrogen volume densities for the three different metallicity scenarios. The symbols are the same as in Fig. \ref{fig:metallicities}.}
\end{figure}

The total hydrogen volume densities computed from the three different metallicity sources are shown in Fig. \ref{fig:m74_ntot}. No significant difference is present, although the values obtained from the SpIOMM metallicity map are consistently higher because of the lower dust-to-gas ratios. The median difference in dust-to-gas ratios for candidate PDRs present in both metallicity maps is 0.38, where the PINGS dust-to-gas ratios are mostly higher. The median difference in total hydrogen volume densities is about 10 $\rm{cm}^{-3}$.
The overall range of densities is between $\sim5 \rm{cm}^{-3}$ and $\sim700 \rm{cm}^{-3}$ with a few higher density outliers.

\begin{figure}
  \resizebox{\columnwidth}{!}{\includegraphics{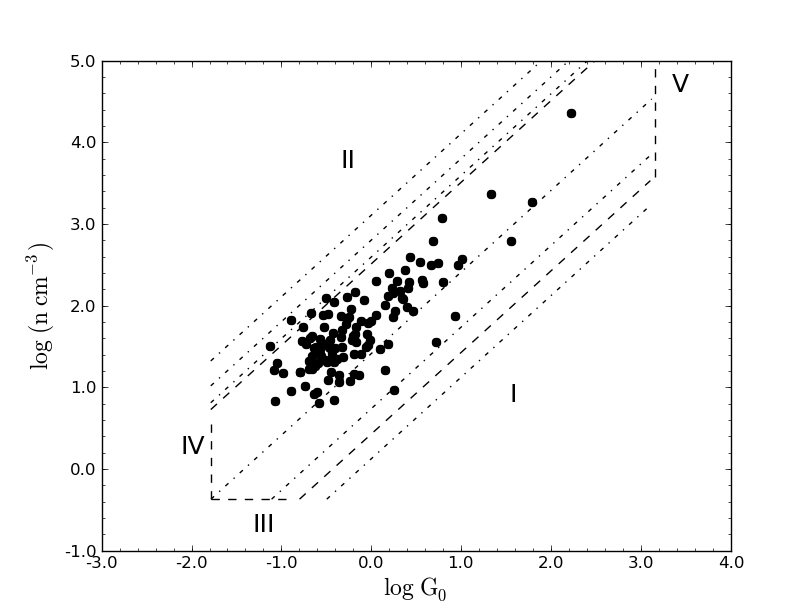}}
  \caption{\label{fig:m74_G0_n} Total hydrogen densities plotted against incident UV flux $G_0$. The different selection effects are shown and described in the text.}
\end{figure}

Finally the plot with the observational selection effects ($G_0$ vs. $n$) is shown in Fig. \ref{fig:m74_G0_n} (only for the results based on the SpIOMM metallicity map). This figure is similar to the ones in \citet{2004ApJ...608..314A}, e.g. their Figure 3, but the axes are switched. As can be seen in their figures, constant HI column densities correspond to straight lines in this type of plot. 
The Roman numerals indicate the selection effects:

\makeatletter
\renewcommand{\theenumi}{\Roman{enumi}}
\renewcommand{\labelenumi}{\theenumi.}
\makeatother
\begin{enumerate}
  \item This is the HI column density upper limit of $5 \times 10^{21}~\rm{cm}^{-2}$ at a typical $\delta/\delta_0$ of 0.6. Theoretically, the HI column becomes optically thick at $5 \times 10^{21}~\rm{cm}^{-2}$, but this value is not observed due to beam smoothing (see H08b). If a value of $3 \times 10^{21}~\rm{cm}^{-2}$ were used, the line would be closer to the point cloud. As \citet{2004ApJ...608..314A} noted, for typical observations the optically thick limit is a few times $10^{21}\ \rm{cm}^{-2}$ depending on spin temperatures, HI profile linewidths and local opacities. We also show the position of this limit for three additional $\delta/\delta_0$ of 0.3, 0.5 and 0.7 (dash dotted lines), where the highest value corresponds to the lower-right line.
  \item The HI lower limit of $2.9 \times 10^{20}~\rm{cm}^{-2}$ at a typical $\delta/\delta_0$ of 0.8 is a sensitivity limit. The dash dotted lines show the same limit using $\delta/\delta_0$ of 0.3, 0.5 and 0.7, where the lowest value corresponds to the upper-left line.
  \item The lowest volume density that we can observe as a consequence of the radio beam diameter and the HI sensitivity limit. For M74, it is approximately $0.4~\rm{cm}^{-3}$. This limit is independent of our method and in practice we don't obtain values this low due to the dust-to-gas ratio being low (relative to e.g. M83).
  \item The minimum $G_0$ ($\sim 0.01$) that we can use is a combination of the lowest measured flux and the largest acceptable value of $\rho_{\rm{HI}}$. In practice a stronger cut-off is the source contrast. The values of $G_0$ drop below one at a value of about log($G_0$) = -1.2, explaining the actual boundary of the points. This is an observational bias, since the source contrast could be improved with more sensitive data.
  \item The maximum $G_0$ ($\sim 1400$) that we would obtain from the data if the highest flux coincided with the smallest measured separation. As with the previous limit, it is dependent on the observational data. A higher value could be achieved with higher resolution data, to measure smaller separations, or if higher UV fluxes were measured. Source contrast is not an issue in this situation. There are not many flux measurements towards the high end because sources this bright are rare, but additionally it is feasible that the highest fluxes cause HI columns too low to be detected. Therefore, higher resolution would need to go together with higher sensitivity in order to be able to obtain points towards this limit.
\end{enumerate}

\section{Benchmarking against previously obtained M83 densities}

To better understand the performance of our automated density analysis, we compute hydrogen densities for M83 using the same input data as were used in H08b (see Table \ref{tab:properties}). Due to certain limitations in our code, we did not use a variable position angle like we did in that paper, but fixed it at 43 degrees. The results are also limited to the extent of the optical disk, making the fixed position angle less of an issue. The full results on which the plots are based are available through the electronic version of this journal.

\begin{figure*}
  \centering
  \resizebox{\columnwidth}{!}{\includegraphics{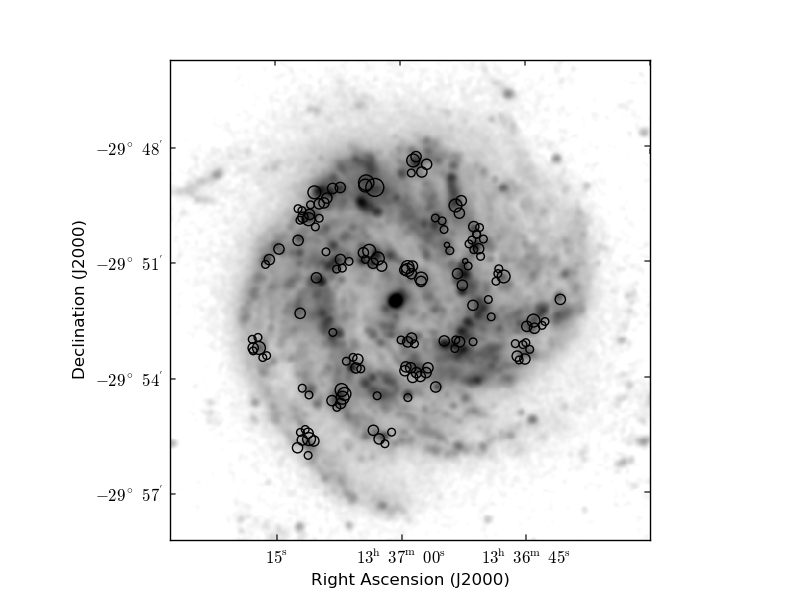}}
  \resizebox{0.75\columnwidth}{!}{\includegraphics{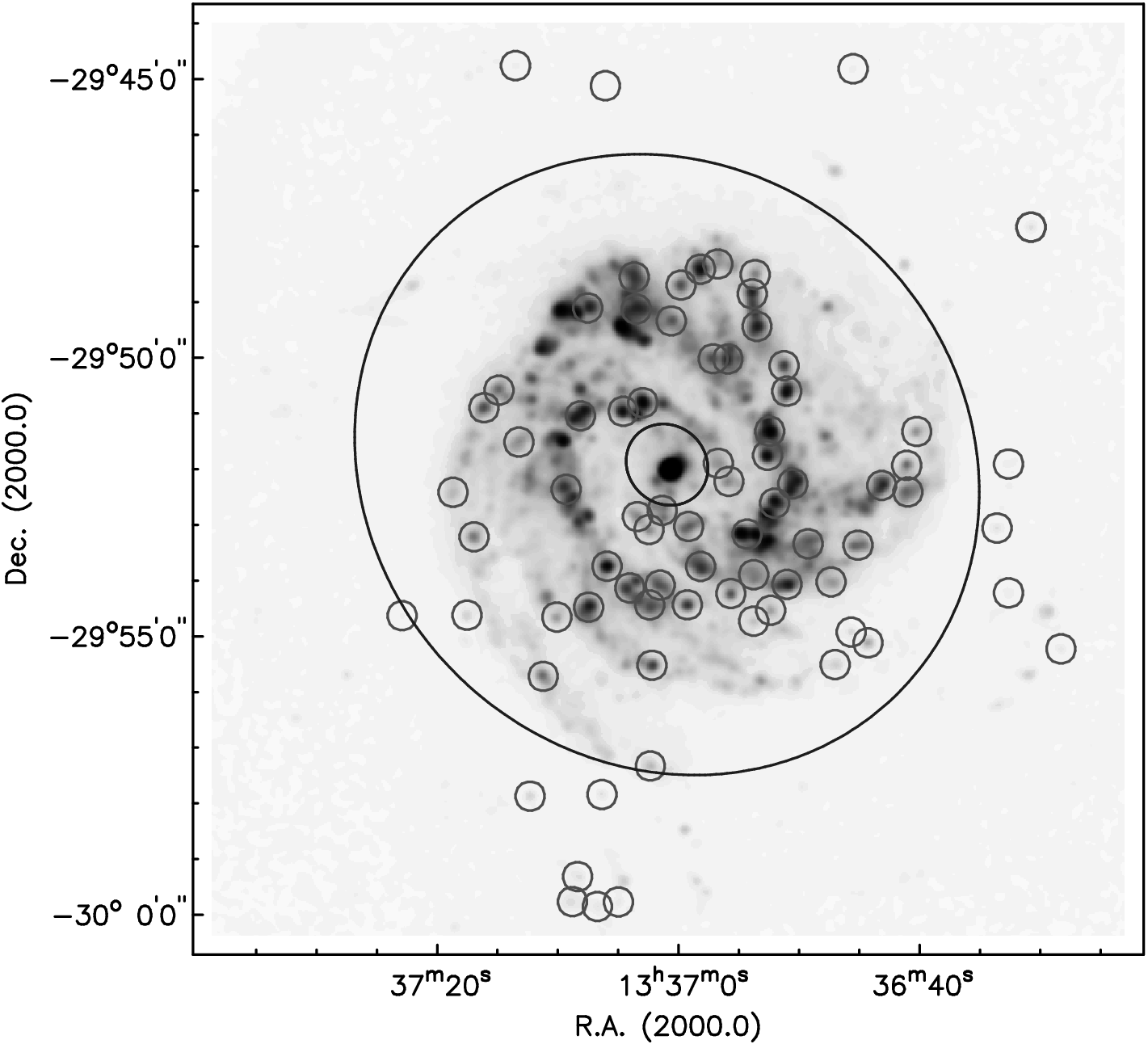}}
  \caption{\label{fig:m83_locplots} Left panel: Total hydrogen volume densities across the disk of M83 near candidate PDRs as identified by clumpfind plotted on top of the GALEX far-UV image. The radius of the circles corresponds to the rounded logarithmic value of $n$. Right panel: Locations of candidate PDRs in M83; Fig. 1 in H08b reproduced.}
\end{figure*}

Fig. \ref{fig:m83_locplots} (left) shows the HI patches surrounding the candidate PDRs located by clumpfind. We reproduce Fig. 1 of H08b showing the candidate PDRs identified in that paper as Fig. \ref{fig:m83_locplots} (right). Note that in the former figure there are multiple circles per candidate PDR, corresponding to individual HI patches that are assumed to be PDR-produced, whereas in the latter figure the central UV sources, or candidate PDRs, are marked.

\begin{figure}
  \resizebox{\columnwidth}{!}{\includegraphics{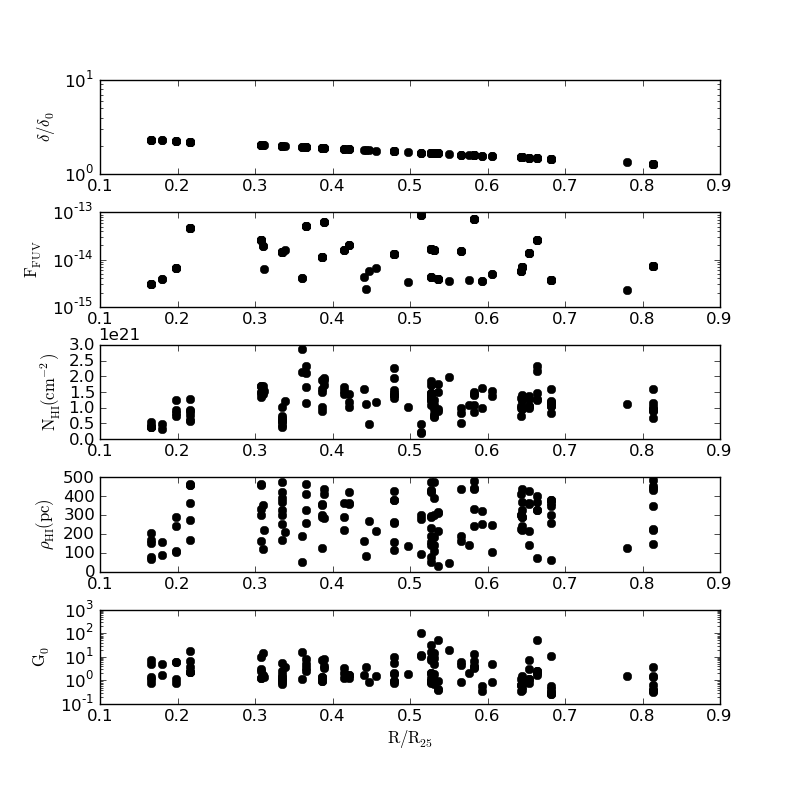}}
  \caption{\label{fig:panel_m83} The measurements and results for M83. From top to bottom: dust-to-gas ratio, far-UV flux (in units of ergs cm$^{-2}$ s$^{-1}$ \AA$^{-1}$), HI column density, separation $\rho_{HI}$ and incident flux $G_0$.}
\end{figure}

Results from the automated PDR method for M83 are shown in Fig. \ref{fig:panel_m83}. From top to bottom: Dust-to-gas ratio, measured far-UV flux, HI column densities, separations between the HI patches and the central UV sources and the incident flux $G_0$ . The dust-to-gas ratio is determined directly from the metallicity gradient. The far-UV fluxes show the usual distribution, diminishing in strength going outwards. The HI column densities as a function of galactocentric radius appear to peak around $\sim0.35~R{25}$, which is slightly sooner than what was observed in H08b, although this cannot be determined very accurately due to the scatter in the column densities. The HI background column density was set to $2.0 \times 10^{20}~\rm{cm}^{-2}$ and subtracted from all measured HI patches, consistent with H08b. The separations between the central source and the HI patches cover the full range of distances (up to 500 pc). This limit was referred to as a 'confusion limit' in H08b, where it was set to 22 arcsec, or almost 500 pc. The results of the automated method were also selected to have a source contrast of at least unity. This is slightly different from H08b, where no such selection was performed, but results with a source contrast below 0.5 were marked. In principle, the source contrast is a measure of how dominant the UV source is with respect to a general background UV radiation field. A value of unity then means that the source is just as bright as the background.

The total hydrogen volume densities obtained here are compared to the ones from H08b in Fig. \ref{fig:m83_R_n}, with the densities from this work in black, and the H08b results in gray (open and closed circles). Only the range of galactocentric radii for which measurements were obtained using the automated PDR method is shown. No significant difference in the range of values can be seen, although on average fewer HI patches were identified by the automated method than by eye in H08b. Since we chose to use only densities from HI patches experiencing a source contrast above 1, this plot shows the H08b results differentiated by this threshold: the open circles are the H08b densities with a source contrast below one. There is no clear separation between the two different sets of densities, although most of the lower-contrast densities have lower density values. Comparing the (logarithmic) densities of the two sets with a source contrast above 1, we find a median value of 1.25 and 1.15 for the H08b densities and the densities obtained here respectively. This corresponds to median densities of 17.6 and 14.1 $\rm{cm}^{-3}$ respectively. Neither set of (logarithmic) densities exhibits a significant correlation coefficient (0.35 and -0.04 respectively), meaning that neither set can be shown to have a correlation between galactocentric radius and (logarithmic) total hydrogen volume densities.

\begin{figure}
  \resizebox{\columnwidth}{!}{\includegraphics{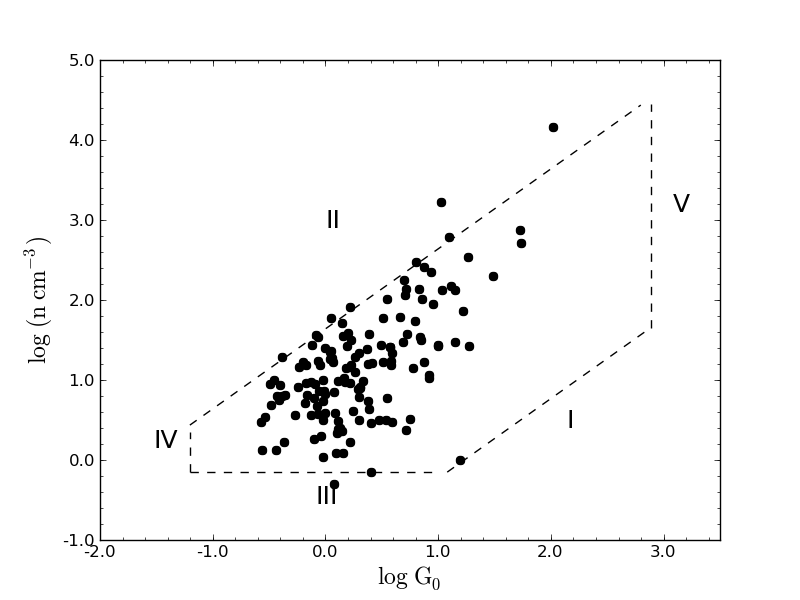}}
  \caption{\label{fig:m83_G0_n} The selection effects affecting the derived M83 total hydrogen volume densities. The individual effects are marked and explained in the text.}
\end{figure}

\begin{figure}
  \resizebox{\columnwidth}{!}{\includegraphics{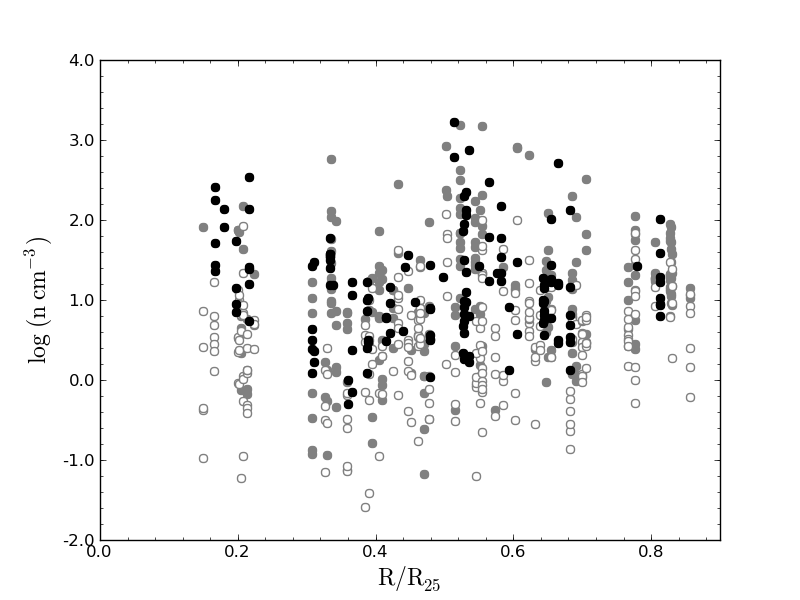}}
  \caption{\label{fig:m83_R_n} Comparison of M83 total hydrogen volume densities from H08b and this work. The gray circles are the H08b densities. Open circles indicate densities obtained from a source contrast lower than 1. The results from this work (black circles) are only those with a source contrast above 1.}
\end{figure}

For completeness, we show how the observational selection effects limit the results in Fig. \ref{fig:m83_G0_n}, with the roman numerals denoting the same limiting factors as in H08b. The limits to the HI column densities are slightly different because of the different range of galactocentric radii covered, and the typical $\delta/\delta_0$ values are different as a result as well.

\makeatletter
\renewcommand{\theenumi}{\Roman{enumi}}
\renewcommand{\labelenumi}{\theenumi.}
\makeatother
\begin{enumerate}
  \item The HI column density upper limit of $3 \times 10^{21}~\rm{cm}^{-2}$ at a typical $\delta/\delta_0$ of 1.9 is appropriate for the range of values obtained here. Theoretically, the HI column becomes optically thick at $5 \times 10^{21}~\rm{cm}^{-2}$, but this value is not observed due to the resolution of the HI map.
  \item The HI lower limit of $3.8 \times 10^{20}~\rm{cm}^{-2}$ at a typical $\delta/\delta_0$ of 2.2 is a sensitivity limit. It is slightly higher than the limit adopted in H08b, since lower HI column densities were identified by eye than automatically by SExtractor. The typical $\delta/\delta_0$ for this value is also different from the one in H08b because in the latter case it was taken at a much larger galactocentric radius.
  \item The lowest volume density that we can observe as a consequence of the radio beam diameter and the HI sensitivity limit. For M83, it is approximately $0.7~\rm{cm}^{-3}$. Due to a slightly higher adopted HI lower limit, this value is slightly higher than the one in H08b. 
  \item The minimum $G_0$ ($\sim0.05$) that we can use. It is higher than the value in H08b (0.005) due to the lowest measured UV flux by the automated method being higher than in H08b. Those lower fluxes originate mostly from larger galactocentric radii (not used here).
  \item The maximum $G_0$ ($\sim770$) that we can obtain from the data. Obtaining this value would require a chance superposition of both the highest UV flux being accompanied by a very close HI patch. It is somewhat higher than the H08b value (565) due to a higher maximum measured flux (since we do not aim to obtain a complete sample of candidate PDRs this can happen), but the actual value would only be obtained by chance.
\end{enumerate}

\section{Discussion and conclusions}

The main factor influencing the uncertainty in the resulting total hydrogen volume density is the dust-to-gas ratio. In principle the use of local metallicity values instead of an averaged slope should improve the accuracy of the results, but this is countered by the intrinsic scatter of the metallicities of individual HII regions. In other words, the individual points should have a smaller error, but the overall result do not change.

It should also be noted that the distance to M74 is far from certain. For example, the distance as quoted in \citet{2011MNRAS.415.2439R} puts M74 a full 2 Mpc further away. However, the effects of this uncertainty on our results are limited: because we use a 'plausibility cut-off' when considering HI patches surrounding candidate PDRs, adopting a larger distance would effectively lower this cut-off, reducing the number of HI patches under consideration. Other measurements are not affected. For example, the value of $G_0$ is dependent on the ratio between $\rho_{HI}$ and the distance to M74. Finally, a different distance could influence the selection of candidate PDRs (since we reject the ones that are too close together), potentially leading to a higher density of selected candidate PDRs.

We reiterate that our method provides spot measurements of total hydrogen volume densities. Due to the nature of the one-dimensional, plane-parallel PDR model, we do not obtain information about cloud morphology, nor does the treatment provide a fraction of molecular hydrogen. Additional assumptions about cloud size would have to be made that are beyond the scope of this work. What we obtain are total hydrogen volume densities of GMCs surrounding clusters of OB stars, probing the interstellar medium of the galactic disk wherever these clusters are present. When we compare the range of our input parameters and hydrogen volume densities to the theoretical perspective provided by \citet{2008ApJ...689..865K}, it seems that our values are reasonable for clouds as they are found in our Galaxy.
Additionally, we find no evidence of changing cloud densities as a function of galactocentric radius.

One of the most fundamental assumptions in our approach is taking the HI patches near clusters of OB stars as produced by photodissociation and, more specifically, that these observed HI columns are connected to the nearby young stars. With our new automated method we can gather the kind of multi-galaxy statistics that will allow us to test this underlying assumption by comparisons with tracers of star formation and photodissociation regions.

The findings presented in this paper can be summarized as follows:

\begin{itemize}
  \item We present an automated procedure to calculate total hydrogen volume densities in nearby spiral galaxies, based on the 'PDR method'. This procedure can replace the previous manual application of the PDR method.
  \item We apply this method using metallicity maps for the first time. As compared to using metallicity slopes, the densities resulting from using metallicity maps are more accurate but not significantly different from the ones obtained using a slope. This implies that in general the densities obtained from the PDR method can be expected to be similar regardless of whether a metallicity slope or map is used, which is key to applying this method to larger samples of galaxies.
  \item We show a new metallicity map of M74 obtained with the SpIOMM instrument at the Mont-M\'{e}gantic telescope and compare it to a similar map from the literature. The SpIOMM map is limited to the [NII]/H$\alpha$ ratio, but fully covers the extent of the M74 optical disk. 
  \item The range of total hydrogen volume densities obtained in M74 is within $5-700~\rm{cm}^{-3}$.
  \item We apply the method to M83 and compare the results to the earlier results of H08b in order to validate the automated method. Within the more limited range of galactocentric radii, the results are not significantly different. We are therefore confident that we can proceed to apply the automated method to a wider sample of galaxies.
\end{itemize}

We add these total hydrogen volume density measurements of M74 to the list of other spiral galaxies for which these measurements have been carried out, namely M33, M81, M83 and M101. With the automated pipeline described in this paper we will be able to do these measurements for a much larger sample of galaxies, using improved metallicity measurements as they become available. We also intend to compare these densities to other tracers of the interstellar medium as well as more global characterizations of galaxies in order to further validate the simple physical assumption of the presence of photodissociated atomic hydrogen on galactic scales. If this assumption is indeed valid, then these measurements provide an important independent check on CO molecular gas measurements, although additional assumptions are needed to compare the two.

\section*{Acknowledgments}
JSH acknowledges support from a CRAQ postdoctoral fellowship (Centre de Recherche en Astronomie du Qu\'{e}bec), where most of this work was carried out. \note{LRN was supported through grants from the Natural Sciences and Engineering Research Council (NSERC) of Canada and the Fonds Qu\'{e}b\'{e}cois de la Recherche sur la Nature et les Technologies (FQRNT) to Carmelle Robert (Universit\'{e} Laval). SpIOMM is funded by grants from NSERC, the Canadian Foundation for Innovation and FQRNT to Laurent Drissen (Universit\'{e} Laval).}
This paper has benefited from constructive comments from Ron Allen, Fran\c{c}oise Combes, \note{Carmelle Robert and Laurent Drissen,} as well as the anonymous referee.

\label{lastpage}
\bibliographystyle{mn2elong} 
\bibliography{references}

\begin{table}
	\centering
	\begin{tabular}{lll}
		\hline
		Parameter & Value \\
		\hline
		Name 						& NGC 628 (M74) & NGC 5236 (M83)\\
		R.A. (J2000.0) 	&  01:36:41.77  &   13:37:00.92 \\
		Dec. (J2000.0)	& +15:47:00.5   &  -29:51:56.7 \\
		Classification  & SA(s)c        & SAB(s)c\\
		Position angle	& 20\textdegree & 43\textdegree\\
		Declination			& 7\textdegree  & 24\textdegree\\
		Distance 				& 7.3 Mpc       & 4.5 Mpc\\
		R$_{25}$				& 11.0 kpc      & 7.63 kpc\\
		E(B-V)					& 0.070 mag     & 0.066 mag\\
  \end{tabular}
  \caption{\label{tab:properties} Basic properties of M74 and M83. Distance, position angle and declination of M74 as adopted in \citet{2010ApJ...714..571W}. The M83 parameters are identical to the ones adopted in \citet{2008A&A...489..533H}. The M74 optical radius $R_{25}$ is 5.2 arcmin as in \citet{1992A&A...253..335K}. Galactic foreground extinctions E(B-V) are taken from \citet{1998ApJ...500..525S}.}
\end{table}

\clearpage 
\begin{table}
\centering
\rotatebox{270}{
\begin{minipage}{0.5\linewidth}
\begin{tabular}{rrrrrrrrrrrrrr}
\hline\hline
Source & R.A. (2000) & Decl. (2000) & Patch R.A. & Patch Decl. & Radius & $F_{FUV}~^a$ & Aperture & $\rho_{HI}$ & $N_{HI}$ & $G_0$ & $G/G_{bg}$ & $n$ & $^b$ \\
& $(^{h m s})$ & $(^{\circ \prime \prime\prime})$ & $(^{h m s})$ & $(^{\circ \prime \prime\prime})$ & (kpc) & & (arcsec) & (pc) & ($\rm{cm^{-2}}$) & & & ($\rm{cm}^{-3}$) & \\
\hline
1a &   01 36 38.72 &   +15 44 21.93 &  01 36 38.254 &   +15 44 32.70 & 5.81 & 1.29e-14 & 25.5 & 452 & 2.27e+21 &   1.80 &  4.81 &     9 & 0.46\\
1b & & &  01 36 39.188 &   +15 44 25.17 & & & & 266 & 2.07e+21 &   5.19 & 13.85 &    36 & 0.48\\
2a &   01 36 45.16 &   +15 47 48.93 &  01 36 45.451 &   +15 47 45.98 & 2.44 & 2.62e-15 & 12.0 & 183 & 9.08e+20 &   2.24 &  9.21 &   120 & 0.50\\
2b & & &  01 36 45.244 &   +15 47 51.99 & & & & 116 & 8.43e+20 &   5.52 & 22.71 &   333 & 0.61\\
2c & & &  01 36 44.413 &   +15 47 55.02 & & & & 441 & 9.17e+20 &   0.38 &  1.58 &    20 & 0.42\\
2d & & &  01 36 45.971 &   +15 47 44.47 & & & & 447 & 5.15e+20 &   0.38 &  1.54 &    47 & 0.50\\
3a &   01 36 57.74 &   +15 47 08.40 &  01 36 57.604 &   +15 47 06.60 & 8.20 & 6.68e-15 & 16.5 &  94 & 1.34e+21 &  21.39 & 99.92 &  2333 & 0.67\\
3b & & &  01 36 57.606 &   +15 47 17.09 & & & & 316 & 1.36e+21 &   1.91 &  8.93 &   203 & 0.39\\
4a &   01 36 44.43 &   +15 44 57.93 &  01 36 44.284 &   +15 45 01.02 & 4.54 & 6.38e-15 & 16.5 & 133 & 1.45e+21 &  10.29 & 36.19 &   369 & 0.53\\
4b & & &  01 36 43.764 &   +15 44 52.04 & & & & 400 & 1.01e+21 &   1.14 &  4.00 &    76 & 0.40\\
5a &   01 36 47.03 &   +15 45 47.43 &  01 36 47.304 &   +15 45 45.93 & 3.73 &  3.7e-15 & 18.0 & 151 & 9.51e+20 &   4.64 & 14.00 &   317 & 0.52\\
5b & & &  01 36 47.617 &   +15 45 54.92 & & & & 401 & 9.41e+20 &   0.66 &  1.98 &    46 & 0.41\\
5c & & &  01 36 46.888 &   +15 45 42.94 & & & & 175 & 7.44e+20 &   3.46 & 10.44 &   342 & 0.51\\
\hline
\multicolumn{14}{l}{$^a$ $\rm{ergs\ cm^{-2}\ s^{-1}\ \mathrm{\r{A}}^{-1}}$}\\
\multicolumn{14}{l}{$^b$ Fractional error}
\end{tabular}
\caption{\label{tab:results} Locations and properties of candidate PDRs (example, full dataset available through the electronic version of this journal).}
\end{minipage}
}
\end{table}


\end{document}